# rECGnition_v1.0: Arrhythmia detection using cardiologist-inspired multi-modal architecture incorporating demographic attributes in ECG

**Shreya Srivastava**
*Indian Institute of Technology Roorkee, Roorkee, India*
ssrivastava@bt.iitr.ac.in

**Durgesh Kumar**
*Indian Institute of Technology Roorkee, Roorkee, India*
dkumar@ce.iitr.ac.in

**Jatin Bedi**
*Thapar Institute of Engineering and Technology, Patiala, Punjab, India*
jatin.bedi@thapar.edu

**Sandeep Seth**
*All India Institute of Medical Sciences, New Delhi, India*
drsandeepseth@hotmail.com

**Deepak Sharma**
*Indian Institute of Technology Roorkee, Roorkee, India*
deepak.sharma@bt.iitr.ac.in

## ABSTRACT

***Objective:*** A substantial amount of variability in ECG manifested due to patient characteristics hinders the adoption of automated analysis algorithms in clinical practice. None of the ECG annotators developed till-date consider the characteristics of the patients in a multi-modal architecture. ***Methods:*** We employed the XGBoost model to analyze the UCI Arrhythmia dataset, linking patient characteristics to ECG morphological changes. The model accurately classified patient gender using discriminative ECG features with 87.75% confidence. We propose a novel multi-modal methodology for ECG analysis and arrhythmia classification that can help defy the variability in ECG related to patient-specific conditions. This deep learning algorithm, named rECGnition_v1.0 (robust ECG abnormality detection version 1), fuses Beat Morphology with Patient Characteristics to create a discriminative feature map that understands the internal correlation between both modalities. A Squeeze and Excitation based Patient characteristic Encoding Network (SEPcEnet) has been introduced, considering the patient's demographics. ***Results:*** The trained model outperformed the various existing algorithms by achieving the overall F1-score of 0.986 for the ten arrhythmia class classification in the MITDB and achieved near-perfect prediction scores of ~0.99 for LBBB, RBBB, Premature ventricular contraction beat, Atrial premature beat and Paced beat. Subsequently, the methodology was validated across INCARTDB, EDB and different class groups of MITDB using transfer learning. The generalizability test provided F1-scores of 0.980, 0.946, 0.977, and 0.980 for INCARTDB, EDB, MITDB AAMI, and MITDB Normal *vs*. Abnormal Classification, respectively. ***Conclusion:*** Therefore, with a more enhanced and comprehensive understanding of the patient being examined and their ECG for diverse CVD manifestations, the proposed rECGnition_v1.0 algorithm paves the way for its deployment in clinics.

# HIGHLIGHTS

- This article documents changes induced by physical parameters into ECG morphology using machine learning and publicly available datasets.
- Apart from good performance and enhanced generalized capability, we propose a novel future-first methodology for understanding and classifying heartbeats.
- Demonstrate that the automated deep learning algorithm, rECGnition_v1.0, can utilize knowledge from various sources to mimic cardiologists' actual ECG analysis mechanism successfully.
- We have extensively evaluated and benchmarked results for four ECG datasets using our multi-modal architecture approach.

# GRAPHICAL ABSTRACT

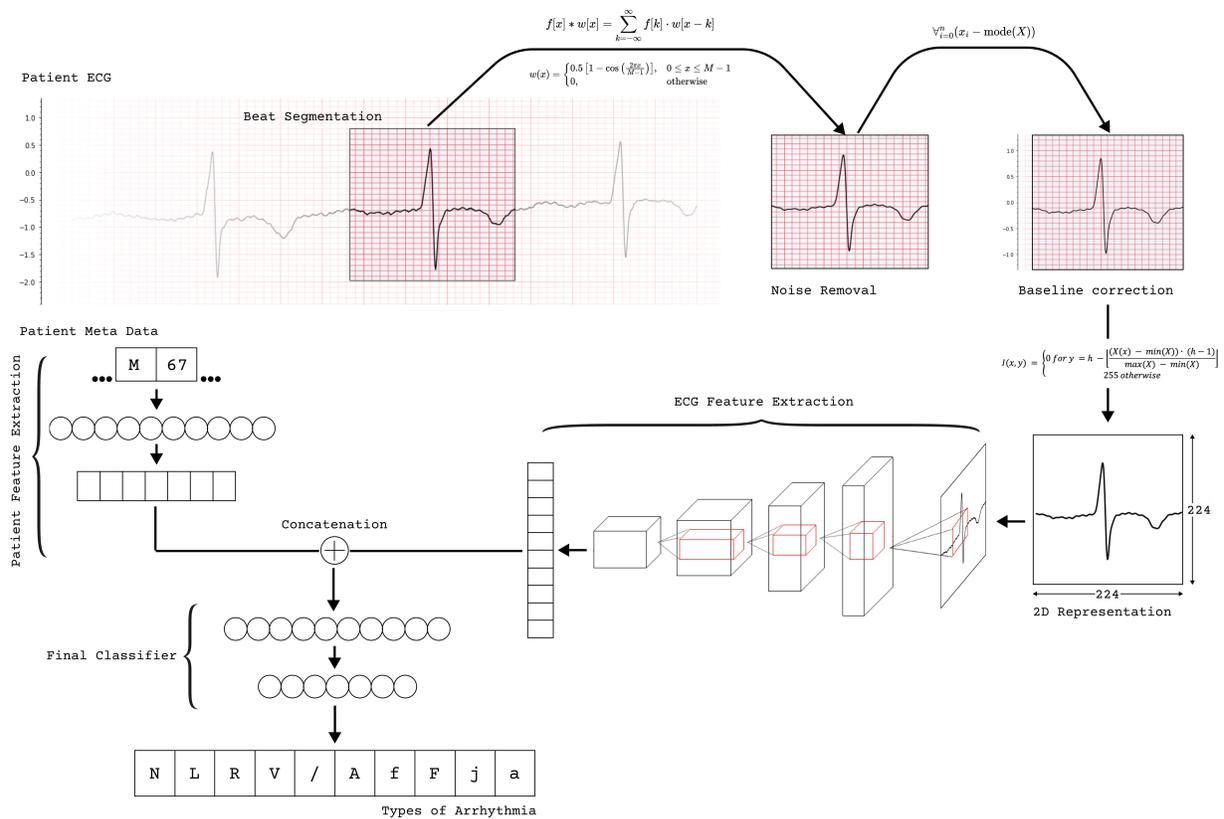

# 1. Introduction

Over the years, cardiovascular diseases (CVDs) have increased significantly, almost doubling from 271 million in 1990 to 523 million in 2019 [1]. In addition, CVD-related deaths have also risen steadily, with numbers going up from 12.1 million in 1990 to 20.5 million in 2021 [2]. This steady rise in cardiovascular diseases poses a diagnostic burden on the healthcare infrastructure, primarily in developing countries like those in Africa, with higher concentrations and rates of increase in CVDs due to the unaffordability of healthcare services [3, 4]. The detection of CVDs has become eminent considering its statistical trends, but it has long been a challenging task, plagued by several hurdles. Traditional diagnostic methods often lack sensitivity and specificity, leading to false-positive (FP) or false-negative (FN) results [5]. Additionally, manual ECG signal interpretation is time-consuming and requires expert knowledge. Hence, in the near future medical informatics will play a pivotal role in improving the healthcare [6-9]. Machine learning and deep learning techniques have shown promise in improving the accuracy of CVD detection by analyzing the ECG signals [10-12]. Various methods have been developed for ECG analysis, like linear discriminator using RR interval based approach for arrhythmia categorization [13, 14], fuzzy Neural Network model and Hermite function for the feature extraction [15, 16], CNN-LSTM Fusion networks for temporal information of ECG signal [17, 18], non-linear decomposition methods and Support Vector Machine (SVM) [19-21], Random Forest (RF) [22, 23], $k$-Nearest Neighbour (kNN) [24, 25], and CNN-based Deep Neural Network [26-28]. These approaches can automate interpretation, reduce human error and enable real-time patient monitoring. Despite achieving near-perfect prediction capability for abnormal heart conditions like arrhythmias, real-life implications of published algorithms remain low. This is due to the inability to generalize over a diverse population as their validation was done on a very restricted and smaller data sample, the failure to correlate patients' characteristics ($P_c$) with the ECG morphology ($E_m$), and the lack of comprehensiveness in the prediction outcomes [29]. In real diagnosis scenarios, an experienced cardiologist considers the multitude of information about the patient being examined along with ECG test results and has a wider array of abnormalities in mind while making diagnostic decisions about the patient's health from the ECG report [30]. Our study improves the real-life medical adoption of automatic ECG analysis algorithm by delivering on the aforementioned problems. We have developed rECGnition_v1.0 (robust ECG abnormality detection version 1) algorithm that simulates the procedural structure followed by cardiologists for inspecting ECG. It fuses the $P_c$ with $E_m$ and builds a correlation map to identify $P_c$-specific patterns, for instance, smaller QRS

---

**Abbreviations:**
ECG: Electrocardiogram; CVD: Cardiovascular diseases; DNN: Deep Neural Network; SVM: Support Vector Machine; RF: Random Forest; CNN: Convolutional Neural Network; LSTM: Long Short-Term Memory; rECGnition_v1.0: robust ECG abnormality detection version 1; SEPcEnet: Squeeze and Excitation based Patient characteristics Encoding network; MITDB: MIT BIH Arrhythmia dataset; INCARTDB: St. Petersburg INCART 12-lead Arrhythmia Database; EDB: European ST-T Database; UCIDB: UCI machine learning repository arrhythmia dataset; $P_c$: Patient characteristics; $E_m$: ECG morphology; TP: True Positive; TN: True Negative; FP: False Positive; FN: False Negative; *N*: Normal beat; *LBBB*: Left Bundle Branch Block beat; *RBBB*: Right Bundle Branch Block beat; *V*: Premature Ventricular Contraction beat; */*: Paced beat; *A*: Atrial Premature beat; *f*: Fusion of paced and normal beat; *F*: Fusion of ventricular and normal beat; *j*: Nodal escape beat; *a*: Aberrated atrial premature beat; AAMI: Association for advancement of mediacl instrumentation; ReLU: Rectified Linear Unit; CDSS: Clinical decision support system.

duration in females than males. SEPcEnet (Squeeze and Excitation based Patient characteristic Encoding network) has been devised to incorporate the patients' demographics. It preserves the raw information and extracted features allowing the meta-classifier direct access to $P_c$. For analyzing $E_m$, we have used CNN based model, which takes an image representation of the heartbeat as input. Given the easy availability of printed ECGs, the shortcoming of analyzing 1D signals like resampling issues, and the error-prone procedure of obtaining digital signals from traditional paper-printed ECGs, it was a calculated decision to use a 2D image representation of $E_m$. UCI machine learning repository arrhythmia dataset (UCIDB) [31] was used to determine $P_c$'s importance in ECG heartbeat classification. Our study then utilized the MIT BIH arrhythmia dataset (MITDB) [32-34] to develop the rECGnition_v1.0 model and benchmark its performance. Subsequently, our algorithm's transferability and applicability across various datasets and data conditions were validated using MITDB (AAMI and Normal *vs.* Abnormal classification), St. Petersburg INCART 12-lead Arrhythmia Database (INCARTDB) [33-35] and European ST-T Database (EDB) [33, 34, 36].

The following are the primary innovations and contributions that this article aims to make:

1. The influence of patient characteristics on ECG has not been well studied in previous deep-learning research despite being an important decision factor in clinical scenarios. This paper documents variation induced by physical parameters to ECG morphology using machine learning and publicly available datasets.

2. The realm of multimodal ECG has not been explored very extensively in the past. This paper offers a novel multimodal methodology that can comprehensively understand ECG signals for distinct patients and improve classification performance and generalization capability.

3. Artificial Intelligence is progressing towards emulating the complexities of real-world processes. This paper demonstrates that an automated Deep Learning (***rECGnition_v1.0***) algorithm incorporating knowledge from various sources can mimic cardiologists' actual ECG analysis mechanism.

4. This paper thoroughly evaluates and benchmarks three ECG datasets (MITBIH, INCARTDB and EDB) by employing a multimodal deep learning architecture tailored for the classification of arrhythmias.

## 2. Literature Survey

Incorporating patient demographic and medical data is extremely valuable when analyzing CVDs. A more thorough comprehension of potential anomalies can be attained by adding patient-specific information, such as age, gender, medical history, genetic variations, medication, and other pertinent aspects. As humans' environmental and social conditions are so rapidly changing [37, 38] that even cardiologists have difficulty segregating disease-induced and external parameter-induced changes, this remains a critical problem for automatic mechanisms as well. From the early-mid 20[th] century, various studies have been carried out across the globe to understand the distinctive pattern in beat morphology. Differences in QRS duration have been reported based on the gender and ethnicity of an individual [39, 40]. Several distinctive features of $E_m$ (Figure 1) vary with $P_c$ (Table 1). ECG maps the heart's electrical activity. It is the combination of various deflections that occur away from the baseline. Various calculations of these deviations/points provide us with the heartbeat's morphological feature, which specifies a particular activity happening during a cardiac cycle and the intensity of those activities. Changes in these deflections might occur due to abnormal heart conditions and also shows signs of inconsistency based on individual patients. Due to this correlation, cardiologists use their experience and patient profiles to take any diagnostic decision.

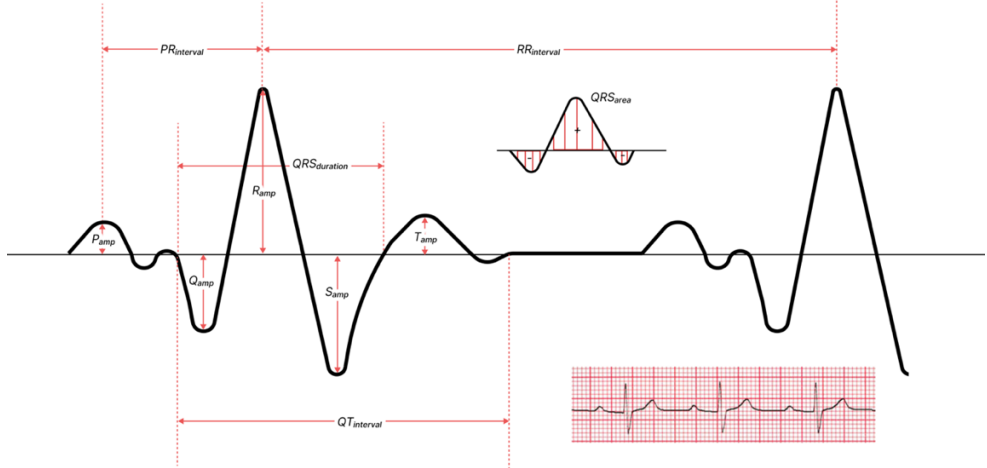

**Figure 1:** The schematic illustration of two heartbeat segments from the ECG signal is shown. Additionally, the figure highlights the diverse morphological feature contained within its construct. Inset shows the cross-section of the printed ECG.

| Physical Parameters | Affected ECG characteristics |
|---|---|
| Ethnicity [41] | Chinese population had the most prolonged PR interval, QRS duration and QTcB interval |
| Age [42] | QRS and T amplitudes decrease and show left axis shift with age |
| Gender [43] | Longer QRS duration in boys than girls, corrected QT interval is longer in females than males |
| Fat (%) [44] | Increased QRS voltage is less prevalent with obesity and increased body fat |

**Table 1**: Physical parameter-induced changes in ECG characteristics.

Doctors and researchers analyze ECG to understand/detect various CVDs. For instance, an abnormally short PR interval on ECG during sinus rhythm results in Wolff-Parkinson-White syndrome [45]. On the other hand, the prolongation of QT intervals causes long QT syndrome [46, 47]. These findings are based on manually analyzing records collected from heart patients, which requires a high level of domain expertise [48]. However, due to the advancement in computational statistics and machine learning, new doors for ECG analysis have opened; researchers have been finding different ways to extract information from ECG [49]. A real-time ECG processing algorithm and a quantitative approach to predict exercised-induced ischemia were developed [50]. Over the past several decades, various methods incorporating adaptive filtering [51], wavelet transformation [52, 53], ANN [54], autoregressive modeling [55], SVM [56], Probabilistic Neural Network [57], STFT-based spectrogram classified using CNN [58, 59], Deep coded feature and LSTM networks [60], LSTM-based autoencoders [61, 62], combined Fuzzy KNN and ANN [63, 64], and RNN [65, 66] have been developed for ECG analysis and arrhythmia detection. Recently, many state-of-the-art papers have been published that use advanced end-to-end deep learning techniques to specifically target heart diseases like arrhythmia, fibrillation, stroke, etc. However, no multi-modal deep learning architecture has been developed that incorporates patient characteristics and ECG data for automatic ECG analysis/classification.

## 3. Material and methods

This study comprises three segments: Firstly, we validated the importance of incorporating the demographic features of the patient for understanding and determining the distinctive changes in heartbeats due to those features using UCIDB. Secondly, we developed the rECGnition_v1.0 algorithm using the conventional MITDB. Towards this end, we prepared and processed the dataset, which involved collecting, cleaning, and transforming the data to make it compatible with the model requirements. Subsequently, we designed the model architecture, which included choosing and configuring neural network layers, optimization algorithms, and regularization techniques. The architecture was carefully designed to incorporate the $P_c$ in ECG signals relevant to arrhythmia classification. Finally, we trained the model using the prepared dataset and evaluated its performance on unseen data. This iterative process involved adjusting hyperparameters, optimizing model weights, and monitoring metrics to enhance the model's accuracy, reliability, and robustness. Thirdly, a comprehensive evaluation was conducted across multiple datasets, including INCARTDB, EDB and various class groups of MITDB, using transfer learning to justify the model's performance and generalization capabilities.

### *3.1 ECG datasets*

The heart's electrical activity was recorded using either two or twelve leads in the ECG datasets utilized in this study (Table 2). The recordings were annotated at both the heartbeat and cardiac rhythm levels, and certain ECGs had been marked explicitly for particular conditions or syndromes (Suppl. Figure S1). The UCIDB dataset is noteworthy because it offered characteristics that were extracted from 12-lead ECG recordings and annotated for the classification of arrhythmias using the extracted feature set. The MITDB dataset was primarily used for the main experimentation and model training. To enhance the assessment of the model's generalizability and usefulness outside of the primary dataset, other datasets like INCARTDB and EDB were used.

| Database | No. of Records | Total Patients | Sample duration | Frequency | Lead count | Total Annotations[#] |
|---|---|---|---|---|---|---|
| MIT-BIH Arrhythmia Database | 48 | 47 | 30m | 360 | 2 | 95948 |
| St.-Petersburg Institute of Cardiological Technics 12-lead Arrhythmia Database | 75 | 32 | 30m | 257 | 12 | 175860 |
| European ST-T database | 90 | 79 | 2h | 250 | 2 | 493253 |
| UCI ML Repository Arrhythmia Dataset | 452 | - | - | - | 12 | 452 |

**Table 2:** Details about different datasets used in our study.
[#]Some records were dropped because of lead mismatch.

## 3.2 Data processing
### 3.2.1 ECG dataset segmentation
The definition of heartbeat used for this study is given by the $Eq.1$ which defines heartbeat as a temporal entity starting from the time instance $(T_{QRSpeak^n} - x)$ and end at time instance $(T_{QRSpeak^n} + x)$. It persists for $2x$ time delta in the continuous vector of potential values of the heart's electrical activity. The occurrence of heartbeat is deterministic and repeats itself after a certain time interval called $RR$ interval. In all ECG datasets used, since the instance of all occurring QRS peaks were annotated, we segmented out the ECG signal $\forall T_n$. The only unknown to the equation $(x)$ was determined based on the sampling frequency of the dataset being examined and the average $RR$ interval duration. For MITDB, the value of $x$ being used is $\sim 280 ms$.

$$(T_{QRSpeak^n} - x) \leq T_n \leq (T_{QRSpeak^n} + x) \tag{1}$$

### 3.2.2 ECG signal pre-processing for standardized image representation
The 2D image representation of the signal is immune to various limitations of 1D signal representation, like non-normalized potential values, irregular sampling frequencies across datasets, etc. Nonetheless, there are some imperfections like uneven baseline activity and noise in potential values of acquired ECG signal, which degrades the quality of sample and negatively affects the model's training. Hence, before translating signal to image, we minimized the noise using a convolution based smoothing technique and corrected the baseline using statistical shift algorithm.

#### 3.2.2.1 Signal smoothening using convolution and baseline correction
We applied Hanning window [67], for smoothening of ECG signal ($Eq.2$ to $Eq.7$). When applied in conjunction with convolution, it effectively reduced spectral leakage and provided better frequency resolution. The segmented signal $v(n)$ was multiplied pointwise with the Hann window function leading to a smoother transition between adjacent segments. The transient part or edge effects at the beginning and the end of the convolved signal $V$ which introduced unwanted artifacts and distortions, were handled using reverse reflection mechanism described by $Eq.5$. Simple edge trimming process was used to obtain the initial length $(n)$ of the signal; the final signal obtained after $Eq.7$ had $N$ more data points.

*Signal:*
$$v(n) = [v, v_1, v_2, \ldots, v_n] \tag{2}$$

*Hanning Window:*
$$w(n) = 0.5 \left(1 - \cos\left(\frac{2\pi n}{N}\right)\right), 0 \leq n \leq N \tag{3}$$

*Discrete Convolution:*
$$(w * v)n = \sum_{m=-\infty}^{\infty} w_m v_{n-m} \tag{4}$$

*Smooth Signal:*
$$v'(n) = [v_N, \ldots, v_{n-N-1}, v_0, v_1, v_2, \ldots, v_n, v_0, \ldots, v_{N-1}] \tag{5}$$
$$w'(n) = [w_0, w_1, w_2, \ldots, w_{N-1}] \tag{6}$$
$$V = (w' * v')(n) \tag{7}$$

$$= \sum_{m=-\infty}^{\infty} [w_0, w_1, w_2, \ldots, w_{N-1}]_m [v_N, \ldots, v_{n-N-1}, v_0, v_1, v_2, \ldots, v_n, v_0, \ldots, v_{N-1}]_{n-m}$$

The window size ($N$) was made by manual inspection of the randomized set created by selecting ~150 beats from all different classes. The noise-free construct was compared with the original construct; amplitude and interval lengths were considered to analyze the overall effectiveness of the selected window. A shift transformation technique was employed to align the baseline of the ECG signal with the X-axis of the plot. This technique uses the mode of the ECG signal to identify the baseline position and shift it to its $0\ mv$ potential value position, enabling consistent and standardized representation across ECG recordings. This alignment ensured that the important features of the ECG waveform, such as P-waves, QRS complexes and T-waves, are accurately represented relative to the baseline.

The standardized representation of the filtered signal in the form of a 2D grid was achieved by restricted axis plotting of the baseline corrected ECG signal. We set the $(min, max)$ range of the plot axis by computing these values $\left( \min_i \{beats_i\}_{i=1}^n\ ,\ \max_i \{beats_i\}_{i=1}^n \right)$ respectively; extreme outliers were tackled beforehand.

*3.2.3 Addressing class imbalance using resampling and augmentation techniques*
Every dataset that we used had a high-class imbalance even for classification into Abnormal *vs.* Normal beats, the class ratio was *1:2.35* and the situation was worse for 10 class classifications with the ratio of *1:500* for Aberrated atrial premature Beat (a): Normal Beat (N). To prevent the model from giving more weightage to some class which has more data points, we utilized two techniques that were loss penalization and dataset resampling [68, 69]. Loss penalization had very flaky results over different runs, so we discarded it and used resampling techniques, yielding consistent and improved results over penalization. Stratified splits of data [70] were created based on age, gender and arrhythmia class. Each split had an almost similar distribution as of the sample dataset; based on arrhythmia count, we randomly discarded splits to down the sample for their respective class. We also tried upsampling certain classes using augmentation techniques. Still, results did not show improvement because only one transformation was practically correct, i.e., horizontal shift, and due to the shift-invariant nature of the CNN.

*3.3 rECGnition_v1.0 architecture and fusion strategy for ECG Analysis*
The rECGnition_v1.0 architecture proposed in this study was developed to effectively classify heartbeat anomalies by considering both the heartbeat image ($X_1$) and P$_c$ ($X_2$) [Figure 2]. The input $X$ consisted of a 2D vector $X_1$ and a 1D vector $X_2$. This combined information played a crucial role in establishing correlations and its mapping to the ground truth labels. The architecture was specifically tailored to holistically process the entire input, $X$. The heartbeat feature extraction unit, referred to as the heartbeat encoder, employed a CNN-based approach inspired by the seminal work of LeCun *et al.* [71]. CNNs have demonstrated exceptional performance in spatial-based classification tasks. To leverage these advancements, we adopted the efficient net [72] model as the backbone for the Heartbeat Feature Extractor network (HbFEnet). This model processed fixed size 224x224 heart images through a series of convolutional layers with different filter sizes, followed by activation functions and pooling layers. This process resulted in extracting high-level features that captured relevant spatial patterns and structures in the heart images. A dedicated network called the squeeze and excitation-based patient characteristics encoder network (SEPcEnet) was devised to incorporate the patient characteristics. SEPcEnet effectively encoded the input $X_2$, capturing the actual information and its correlated features. The feature map of the heartbeat encoder and the output vector from the SEPcEnet were combined using a late fusion approach, represented by $Eq.\,8$. It allowed different models to train for distinct modalities, giving us more flexibility to handle each input as a distinct entity and choosing models that best suits that modality. This fusion facilitated the integration of both the visual information of $E_m$ and the contextual information from the $P_c$. The concatenated vector $H$ served as the input for the meta classifier network (Mcnet), which was a DNN-based component responsible for mapping the combined information to the actual ground truth labels.

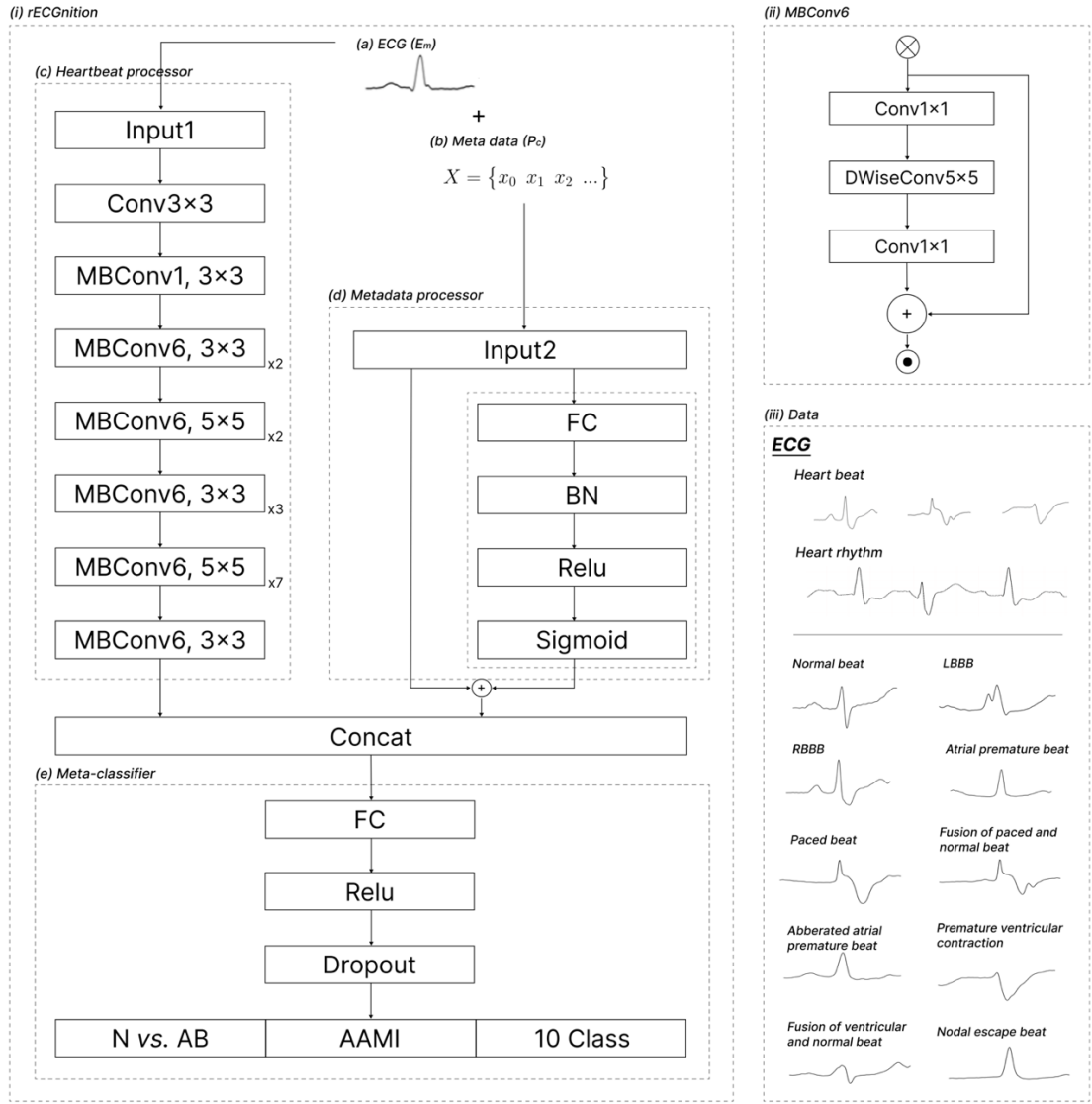

**Figure 2:** (i) The overall architecture design of rECGnition_v1.0 showing (a) ECG [$E_m$], (b) Metadata [$P_c$], (c) Heartbeat processor, (d) Metadata processor, (e) Meta-classifier, (ii) MBConv6, (iii) Data (N: Normal, AB: Abnormal).

Mathematically, the process can be summarized as follows:

*Heartbeat Feature Encoder*: $f_{E_m}(X_1) = \text{HbFEnet}(X_1)$

*Patient Characteristics Encoding*: $f_{P_c}(X_2) = \text{SEPcEnet}(X_2)$

*Late Fusion*: $H = [f_{E_m}(X_1), f_{P_c}(X_2)]$

*Classification*: $y_{pred} = \text{MCnet}(H)$

$$y_{pred} = \text{MCnet}([\text{HbFEnet}(X_1), \text{SEPcEnet}(X_2)]) \tag{8}$$

In the above equations, $f_{E_m}$ denotes the output of the heartbeat encoder, $f_{P_c}$ represents the output of the SEPcEnet, H denotes the concatenated vector of the two modalities, and $y_{pred}$ indicates the predicted probabilities for each class obtained from the DNN-based metaclassifier.

### *3.4 Loss function and optimization strategy*

We combined the Adam optimizer Field [73] and the cosine decay with a linear warm-up learning rate scheduler to minimize cross-entropy loss during model training [74]. The Adam optimizer, which is renowned for its effectiveness in optimizing deep learning models, efficiently alters the model's parameters throughout training. A learning rate scheduler was also utilized to modify the learning rate dynamically. This scheduler decreased the learning rate progressively according to a cosine-shaped decay function (Suppl. Figure S2). The linear warmup phase at the beginning of training helps in stabilizing the learning process by progressively accelerating the learning rate. The final updated parameter is given by the last $Eq. 15$.

$$J = -\sum_{i=1}^{N} y_i \log(\hat{y}_i) \tag{9}$$

$$\nabla_\theta J = \left(\frac{\delta J}{\delta \hat{y}}\right) * \left(\frac{\delta \hat{y}}{\delta \theta}\right) \tag{10}$$

$$m_t = \beta_1 m_{t-1} + (1 - \beta_1)\nabla_\theta J \tag{11}$$

$$v_t = \beta_2 v_{t-1} + (1 - \beta_2)(\nabla_\theta J)^2 \tag{12}$$

$$\alpha(t) = \left(\frac{t}{t_w} * LR_i\right) \; ; t < t_w \tag{13}$$

$$\alpha(t) = max\left(0, \left(1 + cos\left(0.5\pi * \frac{t - t_w}{max(1, T - t_w)}\right)\right) * LR_i\right) \; ; t \geq t_w \tag{14}$$

$$\theta(t) = \theta(t-1) - \alpha(t) * \frac{m_t}{\sqrt{v_t} + \varepsilon} \tag{15}$$

$J$ defines the equation for the cross-entropy loss function, $N$ denotes the number of classes where $\hat{y}_i$ represents the model's prediction and $y_i$ is ground truth. $\beta_1$ and $\beta_2$ are forgetting parameters. $LR_i$ is the learning rate value, $t_w$ and $T$ represents the steps for warmup and total steps, respectively. $\varepsilon$ is a small number typically taken from $10^{-7}$ to $10^{-10}$. Using the above optimization strategy, model converges well as shown in Suppl. Figure S3.

*3.5 Hyperparameter optimization and k-fold validation*
Deep learning relies greatly on hyperparameter optimization, substantially improving models' accuracy and practical applicability. Determining the optimal hyperparameter combination using conventional methods can be exhausting and computationally intensive. Grid Search, Random Search [75] and Bayesian Search are among the popular techniques proposed for determining the optimal hyperparameters for AI models. Bayesian Search [76], which employs probabilistic models to narrow down the search space intelligently, was utilized in this study. Depending on hardware specifications, input size, epochs, learning rate constants and feature extraction backbone, the length of each hyperparameter optimization run varied from 25 minutes to 3.5 hours. Notably, the preponderance of hyperparameter scans was performed only on MITDB-generated dataset splits. The complete process of obtaining the best performing model is represented in Suppl. Figure S4. Further, we utilized $k$-fold training [77], which resamples the training dataset into $k$ groups. In our study, the training dataset was divided into $k = 9$ groups, also known as validation-folds. This approach allowed us to perform separate training on 8 groups while using the remaining group for evaluation purposes (Suppl. Figure S5). Before conducting the $k$-fold training, the stratified splits from the dataset were generated based on age, gender and heartbeat annotation.

*3.6 Performance metrics*
Metrics for measuring the effectiveness of deep learning models in multi-class classification tasks are essential. These metrics provide information about the model's precision, recall, accuracy, and F1-score to assess how well the model performs the task. Several metrics are frequently used for multi-class classification. A fundamental metric for gauging how accurate predictions are made overall is accuracy. Out of all positive predictions for a given class, precision is the percentage of true positive predictions. The ratio of correctly identified positive instances out of all actually positive instances for a given class is calculated as recall, also known as true positive rate or sensitivity. An accurate evaluation of a model's performance is provided by the F1-score, which is a harmonic mean of precision and recall.

$$Accuracy = (TP + TN) / (TP + TN + FP + FN)$$
$$Precision = TP / (TP + FP)$$
$$Recall = TP / (TP + FN)$$
$$F1\ score = 2 * (Precision * Recall) / (Precision + Recall)$$

**4. Results and Discussion**

*4.1 Analysis of UCIDB for ascertaining patient characteristics*
By using the pattern-finding capability of machine learning, we first tested our hypothesis on a dataset having $E_m$ features; to do so, we utilized the UCIDB. 245 Normal beat characteristics were selected from available 452 data points. Feature importance [78] was calculated for $E_m$ features while trying to group patients based on $P_c$.

**Patient characteristics ($P_c$)**   Age, Gender, Height, Weight

**ECG morphological features ($E_m$)**   (QRS, PR, QT, T, P) Interval, Vector Angle, Heart Rate, (Q, R, S, R', S') Wave Average Width, Intrinsic deflections, Wave Amplitude, etc.

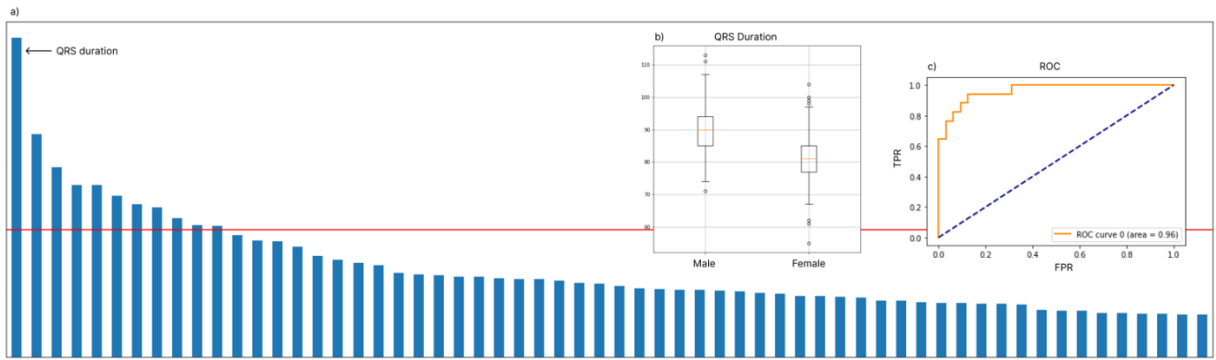

**Figure 3:** Results of analysis conducted on UCIDB. (a) Represents the feature importance of different ECG attributes when finding gender-specific characteristic patterns using XGBoost, (b) Represents the distribution of QRS duration for males and females, (c) ROC-AUC plot for male and female prediction.

While searching for differences in the morphology of males and females, we found out that the QRS interval was a very dominant feature used by the classifier (Figure 3). By closely analyzing the distribution of QRS interval for both groups, it was observed that the average QRS interval for females is slightly smaller than males [79]. The algorithm showed an accuracy of 87.75% for showing gender-based differences (Table 3).

| Parameter | Description | Accuracy |
| --- | --- | --- |
| Gender | For Normal heart conditions predict the gender of the patient based on ECG morphological features (Male/Female) | 87.75% |
| Age | For Normal heart conditions, predict the age of the patient based on ECG morphological features (age <= 45 years / >45 years) | *Females*: 68.75% *Male*: 65.38% *Both*: 67.56% |
| BMI | For Normal heart conditions, predict obesity in patient based on ECG morphological features | 75.67% |

**Table 3:** Result of XGBoost Classifier for ECG Morphology to Patient Metadata association.

The same methodology was then extended to identify age-based and obesity-based differences. Although an accuracy of ~75% was achieved for obesity indicating the effect of BMI on $E_m$, the algorithm exhibited low classification confidence for age-based grouping, probably due to less diversity across age in UCIDB (Table 3).

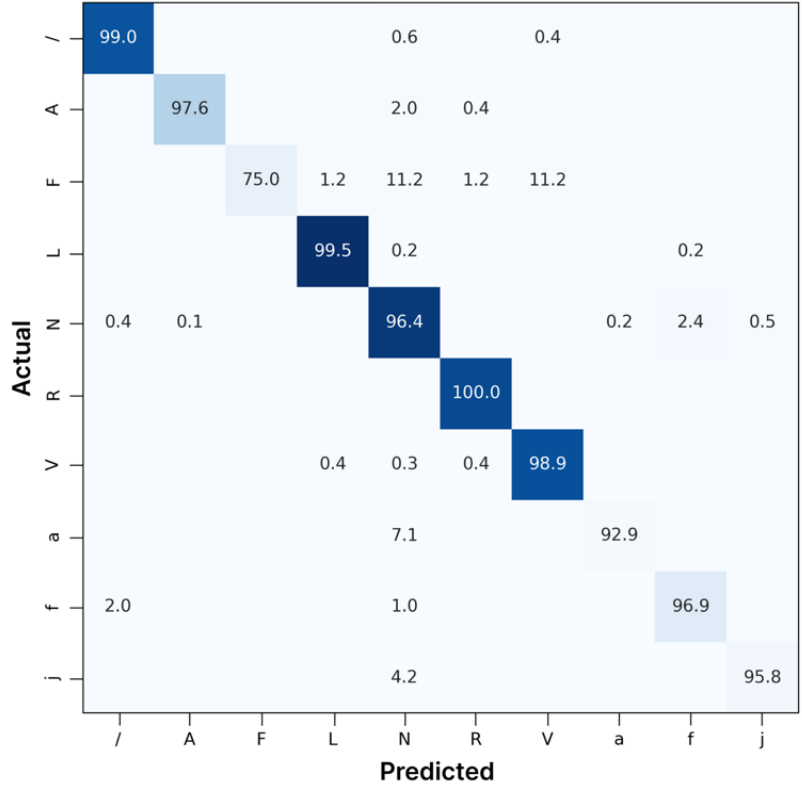

**Figure 4:** Confusion matrix (in %) for heartbeat anomaly detection on MITDB for ten classes using rECGnition_v1.0 architecture.

## *4.2. Performance of rECGnition_v1.0 on MITDB (10 class classification)*

We validated the trained model for classifying 10 different types of heartbeats on MITDB (Experiment $E_0$) [Figure 4, Table 4(a)]. rECGnition_v1.0 achieved an overall F1-score of 0.9855 with a prediction accuracy of 98.56% (Table 5). Among all 10 anomaly classes, LBBB (L) [$Se$: 0.9950; $Pr$: 0.9950] and RBBB (R) [$Se$: 1.00; $Pr$: 0.9932] predictions were off the chart as model displayed a greater understanding of patient-specific changes from arrhythmia induced changes. In addition, high F1-scores of 0.9915, 0.9860 and 0.9861 were obtained for the identification of Paced beat (/), Premature ventricular contraction beat (V) and Atrial premature beat (A), respectively. Given the limited sample size of beats like Fusion of ventricular and normal beat (*F*), Fusion of paced and normal beat (*f*), Nodal (junctional) escape beat (*j*), Aberrated atrial premature beat (*a*), prediction scores were not as compelling as of other beats; despite having such imbalance rECGnition_v1.0 attained an F1-score of >0.85 for all these classes.

## *4.3 Transferability test of rECGnition_v1.0 on INCARTDB, EDB and MITDB (AAMI and N vs. AB)*

Inference from the MITDB was our primary experimental setup ($E_0$). However, to prove the robustness, generalizability and transferability of the model, we performed several experiments ($E_1, E_2, E_3, E_4$ and $E_5$) by changing the datasets and their characteristics. Usually, the deep learning setups follow the strategy of independent train/test splits to build inference metrics; however, the drawback is that the models might not be applicable and transferable to a different/new dataset and therefore lack clinical utility. Hence, it is crucial to make inferences in different dataset conditions.

**Table 4**
Performance of rECGnition_v1.0 on different datasets.

| Classes (10) | Pr | Se | F1 |
|---|---|---|---|
| N | 0.9686 | 0.9637 | 0.9662 |
| L | 0.9950 | 0.9950 | 0.9950 |
| R | 0.9932 | 1.0000 | 0.9966 |
| V | 0.9832 | 0.9888 | 0.9860 |
| / | 0.9929 | 0.9901 | 0.9915 |
| A | 0.9960 | 0.9764 | 0.9861 |
| f | 0.8190 | 0.9694 | 0.8879 |
| F | 1.0000 | 0.7500 | 0.8571 |
| j | 0.8519 | 0.9583 | 0.9020 |
| a | 0.8667 | 0.9286 | 0.8966 |

(a) MITDB for 10-beat classification [*Experiment $E_0$*]

| Classes (2) | Pr | Se | F1 |
|---|---|---|---|
| N | 0.9773 | 0.9877 | 0.9825 |
| AB | 0.9843 | 0.9712 | 0.9777 |

(c) MITDB for Normal *vs.* Abnormal classification [*Experiment $E_5$*]

*N*: Normal beat; *L*: Left Bundle Branch Block beat; *R*: Right Bundle Branch Block beat; *V*: Premature Ventricular Contraction beat; */*: Paced beat; *A*: Atrial Premature beat; *f*: Fusion of paced and normal beat; *F*: Fusion of ventricular and normal beat; *j*: Nodal escape beat; *a*: Aberrated atrial premature beat, *N*: Normal beat; *AB*: Abnormal beat, *SEB*: Supraventricular ectopic beat; *VEB*: Ventricular ectopic beat

| Dataset/Lead | Normal Beat | | | SEB | | | VEB | | | Fusion beat | | |
|---|---|---|---|---|---|---|---|---|---|---|---|---|
| | Pr | Se | F1 | Pr | Se | F1 | Pr | Se | F1 | Pr | Se | F1 |
| **MITDB $E_1$** | 0.9722 | 0.9886 | 0.9803 | 1.0000 | 0.9590 | 0.9790 | 0.9871 | 0.9529 | 0.9697 | 1.0000 | 0.6125 | 0.7597 |
| **INCART DB (V1) $E_2$** | 0.9907 | 0.9972 | 0.9939 | 0.9551 | 0.7526 | 0.8419 | 0.9878 | 0.9868 | 0.9873 | 0.8462 | 0.5410 | 0.6600 |
| **Euro ST-T DB $E_3$** | 0.9484 | 0.9927 | 0.9700 | 0.9286 | 0.5652 | 0.7027 | 0.9731 | 0.9693 | 0.9712 | 0.9388 | 0.5823 | 0.7188 |
| **INCART DB (II) $E_4$** | 0.9900 | 0.9970 | 0.9935 | 0.9818 | 0.7859 | 0.8730 | 0.9839 | 0.9796 | 0.9818 | 0.8718 | 0.5574 | 0.6800 |
| $\Delta(E_2 - E_4)$ [#] | 0.0007 | 0.0002 | 0.0004 | 0.0267 | 0.0333 | 0.0311 | 0.0039 | 0.0072 | 0.0055 | 0.0256 | 0.0164 | 0.0200 |

(b) INCARTDB, EDB, and MITDB for AAMI class classification

[#]$\Delta(E_2 - E_4)$: Represents the difference between results of INCARTDB experiment $E_2$ which was performed on ECG lead V1 and $E_4$ which was performed on lead II; smaller values signify the model's consistency across different Em's.

Consequently, we carried out the transferability test for diverse experimental conditions to demonstrate the utility of rECGnition_v1.0 in clinical practice.

*Experiment 1($E_1$) [MITDB AAMI classification]*: In the context of MITDB AAMI classification, our model achieved an impressive overall accuracy of 97.75%, with an accompanying F1-score of 0.9767 (Table 5). This signifies the robustness of our model's long-term understanding capabilities, which helps it to adapt to new experimental scenarios. Notably, when specifically classifying Fusion beats, our model showed a perfect precision of 1.00 [Table 4(b)], similar to the output of 10 class classifications. This demonstrates the model's ability to transfer knowledge to diverse situations effectively. Furthermore, the F1-scores for Supraventricular ectopic beats (SEB) and Ventricular ectopic beats (VEB) were 0.9790 and 0.9697, respectively. Remarkably, rECGnition_v1.0 exhibited higher precision than sensitivity in these cases, indicating its capability to correctly identify SEB (*Pr*: 1.00; *Se*: 0.9590) and VEB (*Pr*: 0.9871; *Se*: 0.9529) types.

*Experiment 2($E_2$) [INCARTDB AAMI classification using V1 lead]*: In the realm of INCARTDB AAMI classification from V1 lead, rECGnition_v1.0 achieved an outstanding overall accuracy of 98.68%, accompanied by an impressive F1-score of 0.9801 (Table 5). These remarkable results showcase the model's understanding and adaptability across diverse dataset distributions. Notably, when specifically identifying SEB, VEB and Fusion Beat, our model exhibited remarkably high precision rates of 0.9551, 0.9878 and 0.8462, respectively, than sensitivity [Table 4(b)]. Furthermore, our analysis revealed that the model's performance aligns with the findings of MITDB AAMI classification, wherein SEB, VEB and Fusion Beat also demonstrated higher precision values compared to sensitivity. In contrast, Normal Beat exhibited slightly higher sensitivity than precision (*Se*: 0.9972; *Pr*: 0.9907), which too was consistent with the results obtained from MITDB AAMI (*Se*: 0.9886; *Pr*: 0.9722). These findings validated the model's consistency and its remarkable precision in AAMI classification tasks.

*Experiment 3($E_3$) [EDB AAMI classification]*: For EDB AAMI classification, our model achieved an overall accuracy of 95.41% (Table 5), which is slightly lower than the other experimental setups. This can be attributed to the limited number of trainable data points available in the training dataset; for instance, the Normal Beat class consists of 4500 data points, while VEB, SEB, and Fusion Beat have 3096, 291, and 256 data points, respectively. The later numbers are relatively low, which poses a challenge for deep learning models to reach their full potential in terms of performance. Despite this limitation, our model still demonstrated a commendable accuracy in classifying the ECG signals in the EDB for N and VEB, wherein F1-scores were 0.970 and 0.9712, respectively [Table 4(b)].

*Experiment 4($E_4$) [INCARTDB AAMI classification using II lead]*: In this experimental setup, we aimed to evaluate our model's ability to comprehend a distinct heartbeat structure compared to the one it was originally trained on. In comparison to $E_2$, the main/only difference in this experiment was the utilization of a different lead. Remarkably, the obtained results in $E_4$ exhibited a strong correlation with those of $E_2$, with minor variations observed in the prediction of SEB; $\Delta(E_2 - E_4)$ exhibits the prediction discrepancy between $E_2$ and $E_4$ [Table 4(b)]. These findings validated the model's consistent performance and its capacity to generalize its understanding to diverse heartbeat constructs.

*Experiment 5($E_5$) [MITDB Normal vs. Abnormal classification]*: This experiment classified abnormal heartbeats using the MITDB dataset [Table 4(c)]. rECGnition_v1.0 achieved exceptional results, with F1-score of 0.9825 and 0.9777 for the Normal and Abnormal classes, respectively. The overall accuracy achieved was an impressive 98.04% (Table 5).

*4.4 Comparative Study*
The objective of our study was not limited to attaining marginal enhancements in test scores on the test dataset. Instead, we focused on introducing a cutting-edge approach that establishes a framework for advancing CVD prediction by utilizing Artificial Intelligence (AI). To achieve our objective, we meticulously curated the most optimal models available till-date and compared them with rECGnition_v1.0 algorithm (Table 5). Our main aim was to offer a comprehensive comparison of the proposed methodology and its possible ramifications in the field of CVD detection instead of concentrating solely on quantitative metrics. It is acknowledged that substantially feature-engineered and heavily tuned algorithms have demonstrated remarkable outcomes. For instance, Houssein *et al.* [80] reported an overall accuracy of 99.33% and an F1-score of 0.9865 due to extensive feature engineering and hyperparameter tuning for demonstrating their improved Marine Predator algorithm's hyperparameter search capabilities. In contrast, rECGnition_v1.0 was only optimized for 10-class MITDB classification and transferred to other datasets [Table 4(b)]; nonetheless, it outperformed several previously developed algorithms (Table 5).

**Table 5**
Comparisons of the proposed rECGnition_v1.0 with other methods.

| Database | Publications | Classifier | Classes | Performance | | | |
|---|---|---|---|---|---|---|---|
| | | | | Acc | Pr | Se | F1 |
| MITDB | Melgani et al. [81] | SVM | 6 | 91.67 | - | 0.9383 | - |
| | Dutta et al. [82] | LS-SVM | 3 | 95.82 | 0.9701 | 0.8616 | 0.91 |
| | Ince et al. [83] | 1-D CNN | 5 | 96.40 | 0.7920 | 0.6880 | 0.7302 |
| | Jun et al. [84] | 2D CNN | 8 | 98.85 | 0.9859 | 0.9708 | 0.9712 |
| | Mathunjwa et al. [85] | CNN | 3 | 97.21 | 0.9554 | 0.9649 | 0.9596 |
| | Houssein et al. [80] | IMPA-CNN | 4 | 99.33 | 0.9879 | 0.9852 | 0.9865[#] |
| | **Proposed** | **rECGnition_v1.0** | **2** | **98.04** | **0.9804** | **0.9804** | **0.9804**[**] |
| | **Proposed** | **rECGnition_v1.0** | **5** | **97.75** | **0.9779** | **0.9775** | **0.9767**[**] |
| | **Proposed** | **rECGnition_v1.0** | **10** | **98.56** | **0.9866** | **0.9856** | **0.9855**[*] |
| INCARTDB (V1) | Liu Y et al. [86] | CNN + Per patient training | 5 | 97.11 | - | - | - |
| | Houssein et al. [80] | IMPA-CNN | 4 | 99.43 | 0.9890 | 0.9983 | 0.9886[#] |
| | Wang, G et al. [87] | Domain adoption | 2 | 95.36 | - | - | - |
| | V. Kalidas et al. [88] | AE + RF | 2 | NA | 0.9476 | 0.8808 | 0.9130 |
| | Chen G et al. [89] | feature fusion, cascaded classifier | 4 | - | 0.9980 | 0.9980 | 0.9980[#] |
| | **Proposed** | **rECGnition_v1.0** | **4** | **98.68** | **0.9798** | **0.9868** | **0.9801**[**] |
| EDB | Houssein et al. [80] | IMPA-CNN | 4 | 99.75 | 0.9985 | 0.9947 | 0.9951[#] |
| | Jiang et al. [90] | Multi-module | 2 | 93.70 | - | - | - |
| | Krasteva, V. et al. [91] | - | 2 | - | 0.8115 | 0.9649 | - |
| | **Proposed** | **rECGnition_v1.0** | **4** | **95.41** | **0.9404** | **0.9541** | **0.9457**[**] |

[#]Shows slightly better F1-scores than rECGnition_v1.0 as the models either used fine-graded handcrafted features or were extensively hyper-tuned.
[*]rECGnition_v1.0 was developed/trained for classifying 10 classes.
[**]Transferability test showing rECGnition_v1.0's generalizing capability, with high accuracy and F1-scores on new/diverse datasets.

rECGnition_v1.0 demonstrated superior outcomes compared to previously established techniques for beat annotations as it successfully correlated $P_c$ with $E_m$. In RBBB, one of the characterizing factors is widened QRS complex, which overlaps with systematic changes in QRS duration based on the gender of the patient. Due to awareness about the gender of the patient, the model developed the ability to distinguish between the RBBB-widened QRS complex and the already widened male QRS complex. Consequently, rECGnition_v1.0 achieved an F1-score of 0.9966, with a sensitivity of 1.00 towards RBBB classification [Table 4(a)]. Likewise, in the case of LBBB also, there are observed changes in QRS duration; hence, the proposed model exhibited remarkable results for LBBB classification as well.

Translational research is a highly impactful endeavour that consistently aims to generate benefits for society's betterment and prosperity. Complex solutions that involve extensive feature engineering may result in inaccurate predictions due to minor errors in the feature extraction processes. Notably, our approach does not involve advanced feature engineering techniques, thereby conferring a significant general applicability and application readiness advantage. Moreover, since the rECGnition_v1.0 algorithm is reproducible, robust and quite accurate, it can be easily deployed in clinical practice.

## 5. Conclusion

Our study employed the MITDB dataset and sought to improve ECG analysis and the precision of arrhythmia categorization by integrating patient characteristics variables and segmented ECG heartbeats. The rECGnition_v1.0 algorithm demonstrated a noteworthy overall accuracy of 98.56% and an F1-score of 0.9855 across 10 distinct heartbeat categories. Additionally, the model achieved a classification accuracy of 97.75% in AMMI-based beat categorization and 98.04% in distinguishing between Normal *vs.* Abnormal beats on MITDB. The aforementioned results demonstrate the pre-eminence of rECGnition_v1.0 in categorizing anomalous ECG beats. Thus, utilizing a multi-modal approach that integrates $P_c$ in conjunction with $E_m$ represents a more pragmatic and efficacious strategy for classifying arrhythmias. Incorporating $P_c$ helped rECGnition_v1.0 capture vital information essential to deliver precise ECG diagnosis. Integrating individualized $P_c$ and a thorough examination of ECG beats are instrumental in yielding precise and authentic outcomes. To summarize, the present investigation effectively developed and assessed a multi-modal deep neural algorithm, rECGnition_v1.0, for classifying heartbeats, showcasing its superiority over prior models.

## 6. Summary Table

*What was already known on the topic:*
- Deep learning architectures are very effective when it comes to finding correlations among different modalities, and hence, by their incorporation, automatic diagnostic methodologies have achieved specialist-level accuracy.
- Patient demographics and physical parameters are known to influence physiological parameters to the extent that ECG can also be used to identify individuals uniquely.

*What this study added to our knowledge:*
- This work uses machine learning and public datasets to document physical parameter-induced ECG morphological variation.
- Our unique multimodal technique (rECGnition_v1.0) can fully interpret ECG data for different patients and improve classification and generalization.

## CRediT author statement

**Shreya Srivastava:** Conceptualization, Methodology, Software, Validation, Formal analysis, Investigation, Data curation, Visualization, Writing - original draft. **Durgesh Kumar:** Software, Validation, Formal analysis, Investigation, Data curation, Visualization, Writing - original draft. **Jatin Bedi:** Formal analysis, Writing - reviewing and editing. **Sandeep Seth:** Conceptualization, Writing - reviewing and editing. **Deepak Sharma:** Conceptualization, Data curation, Visualization, Writing - reviewing and editing, Resources, Project administration, Funding acquisition, Supervision.

## Acknowledgment

D.S. acknowledges the financial support from MHRD (BT/2014-15/Plan/P-955 and BIO/FIG/100700), SERB (ECR/2016/001566), DBT (BT/PR40141/BTIS/137/16/2021) and DHR (R.11013/51/2021-GIA/HR). S.S. is thankful to MHRD for research fellowship.

**APPENDIX**

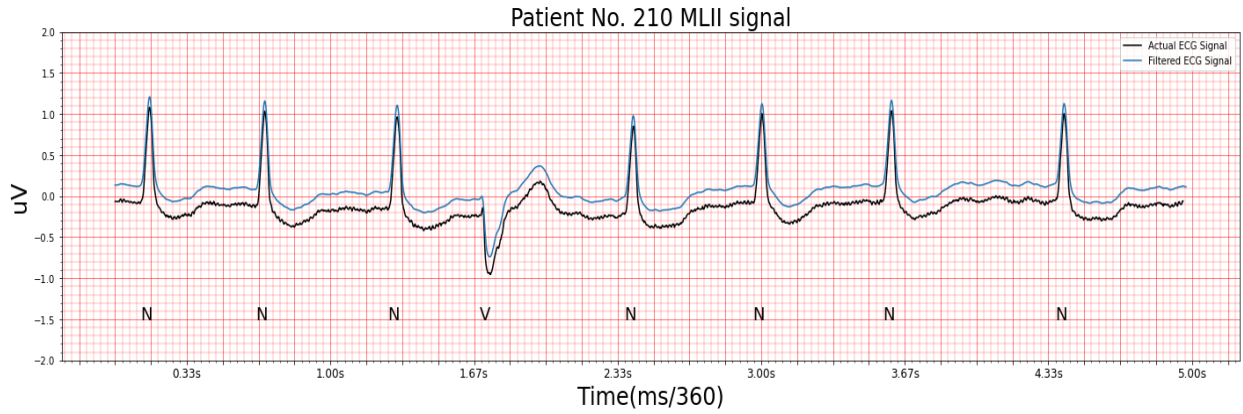

**Suppl. Figure S1:** Representation of ~5-sec recording of patient no. 210 (84-year-old male) from the MITDB.

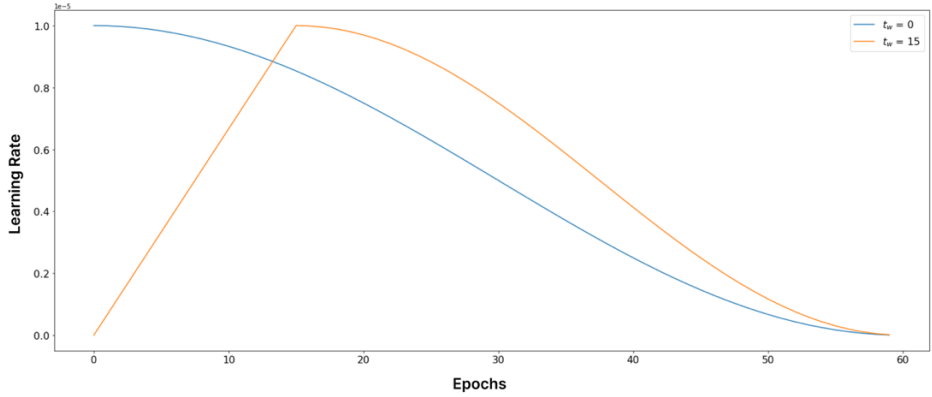

**Suppl. Figure S2:** Cosine decay with linear warm-up learning rate scheduler.

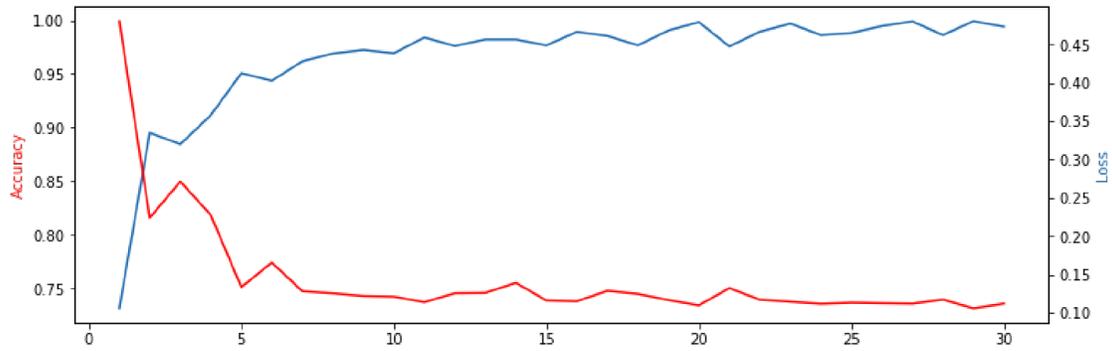

**Suppl. Figure S3:** Training convergence curve for the model

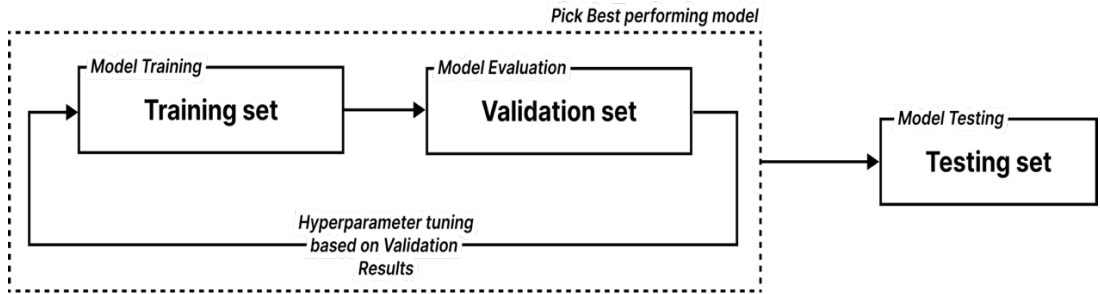

**Suppl. Figure S4:** Flow diagram of Train-Validation-Test loop, used to pick the best possible model and avoid overfitting.

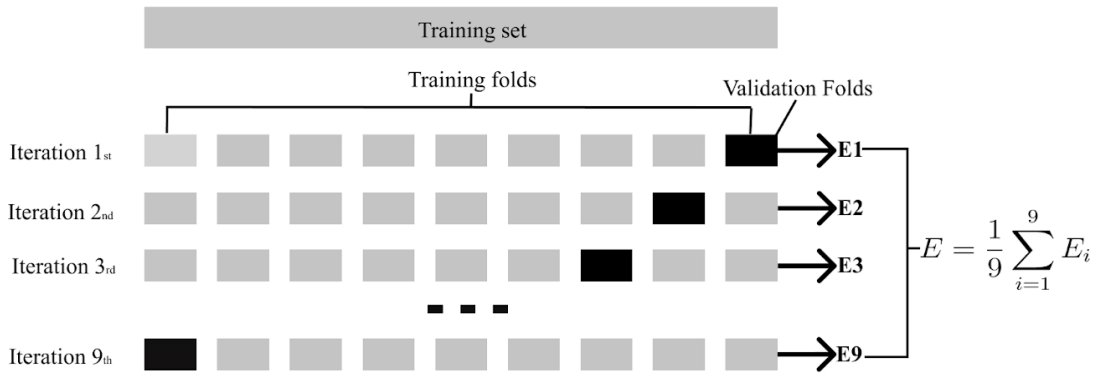

**Suppl. Figure S5:** Diagrammatic representation showing *k*-fold cross-validation for our proposed model.